\documentclass[prd,twocolumn,nofootinbib,superscriptaddress]{revtex4}
\usepackage{amsmath,amssymb,epsfig,bm,mathrsfs,feynmp,slashed,color,graphicx,mathtools}
\usepackage[colorlinks=true,linkcolor=blue,citecolor=blue,urlcolor=blue]{hyperref}
\usepackage{soul}

\newcommand{\bea}{\begin{eqnarray}}
\newcommand{\eea}{\end{eqnarray}}

\newcommand{\vectau}{{\bm \tau}}
\newcommand{\vecrho}{{\bm \rho}}

\def\XXint#1#2#3{{\setbox0=\hbox{$#1{#2#3}{\int}$}
     \vcenter{\hbox{$#2#3$}}\kern-.5\wd0}}

\definecolor{red}{rgb}{0.8,0,0}
\definecolor{violet}{rgb}{0.4,0,0.4}
\definecolor{green}{rgb}{0,0.5,0.0}
\definecolor{navy}{rgb}{0.0,0.0,0.6}
\definecolor{orange}{rgb}{0.8,0.2,0.0}

\usepackage[normalem]{ulem}  

\definecolor{red}{rgb}{0.8,0,0}
\definecolor{violet}{rgb}{0.4,0,0.4}
\definecolor{green}{rgb}{0,0.5,0.0}
\definecolor{navy}{rgb}{0.0,0.0,0.6}
\definecolor{orange}{rgb}{0.8,0.2,0.0}

\begin{document}
\title{
Confronting GW190814 with hyperonization in dense matter and hypernuclear compact stars
}

\author{Armen Sedrakian} 
\email{sedrakian@fias.uni-frankfurt.de } 
\affiliation{Frankfurt Institute for Advanced Studies, D-60438 
  Frankfurt-Main, Germany } 
\affiliation{Institute of Theoretical Physics, University of Wroc\l{}aw,
50-204 Wroc\l{}aw, Poland   
}
\email{sedrakian@fias.uni-frankfurt.de }

\author{Fridolin Weber} 
\affiliation{Department of Physics, San Diego State University, San Diego, California 92182, USA }
\affiliation{Center for
  Astrophysics and Space Sciences, University of California, San
  Diego, La Jolla, California 92093, USA}
\email{fweber@sdsu.edu, fweber@ucsd.edu}

\author{Jia Jie Li}
\affiliation{Institute for Theoretical Physics,
J. W. Goethe University,
D-60438 Frankfurt am Main, Germany}

\begin{abstract}
  We examine the possibility that the light companion in the highly
  asymmetric binary compact object coalescence event GW190814 is a
  hypernuclear star. We use density functional theory with functionals
  that have been tuned to the properties of $\Lambda$ hypernuclei as
  well as astrophysical constraints placed by the masses of the most
  massive millisecond pulsars, the mass-radius range inferred from the
  NICER experiment, and the binary neutron star merger event
  GW170817. We compute general-relativistic static and maximally
  rotating Keplerian configurations of purely nucleonic and
  hypernuclear stars. We find that while nucleonic stars are broadly
  consistent with a neutron star being involved in GW190814, this
  would imply no new degrees of freedom in the dense matter up to 6.5
  times the nuclear saturation density. Allowing for hyperonization of
  dense matter, we find that the maximal masses of hypernuclear stars,
  even for maximal rapidly rotating configurations, are inconsistent
  with a stellar nature interpretation of the light companion in
  GW190814, implying that this event involved two black holes rather
  than a neutron star and a black hole.
\end{abstract}

\date{August 12, 2020}

\maketitle

\section{Introduction}

The recent measurement by the LIGO--Virgo Collaboration (hereafter
LVC)~\cite{Abbott:2020khf} of gravitational waves from a binary
coalescence of a $24.3 M_{\odot}$ black hole with a compact object in
the mass the range {of} $2.50-2.67M_{\odot}$ has raised interest in
the question of whether the light companion is the heaviest known
neutron star (NS) or the lightest known black hole (BH). The mass of
this object falls into the so-called ``mass gap'' {
  $2.5\le M/M_{\odot}\le 5$} where neither a neutron star nor a black
hole have been observed so far.  This observation poses yet another
challenge to the theoretical models of dense nuclear matter in the
light of gravitational wave observations; for reviews see
\cite{Baiotti2019PrPNP,Chatziioannou2020}.

Theoretically, the modern density functional models, which are
compatible with the nuclear phenomenology in the vicinity of
 {nuclear} saturation density  {and} satisfy the mass and/or
radius constraints coming from measurements of massive
($\sim 2.0M_{\odot}$) pulsars~\cite{Cromartie:2019kug},  {the} NICER
experiment~\cite{Miller:2019cac,Riley:2019yda}, and the multimessenger
GW170817 event~\cite{LIGO_Virgo2017a,LIGO_Virgo2017b,LIGO_Virgo2017c},
predict maximum masses of static neutron star sequences of the order
of $2.50M_{\odot}$. As is well known, the maximum mass of a rapidly
rotating NS is  {around} $20\%$ larger than its static
counterpart~\cite{Weber1992,CST94b,Bonazzola1998,Paschalidis:2016vmz}.  {Therefore}
the tension between the NS interpretation of the light companion in
GW190814 and  {the} maximal masses predicted by nuclear models is
mitigated if the  { light companion were} a rapidly rotating  {
  NS}. At the same time, the BH origin of a compact object within the
mass gap range is possible through the prior coalescence of  {a}
binary NS or an NS-white dwarf system, which would suggest that
GW190814 originates from a triple star system.  {Ideas have already
  been put forward regarding the nature of the light companion of
  GW190814 which invoke (stationary or rapidly rotating) very massive
  NS composed of nucleonic
  matter~\cite{Most2020,Fattoyev2020,Tsokaros2020,Zhang2020,Tews2020}
  at the instance of coalescence or in the past.  }

In this paper we examine the possibility  that the
lighter companion in GW190814 is a hypernuclear star and
show that hyperonization in dense matter is incompatible with its
interpretation as a NS, even if one considers a maximally fast
rotating (Keplerian) hypernuclear star. Thus we conclude that the
light companion in GW190814 is a black hole  {if hyperonization
  takes place in compact stars.}

The hyperonization of dense matter is based on a robust energetic
argument which asserts that hyperons will nucleate in dense
 nuclear matter once the neutron Fermi energy reaches the
(in-medium) rest masses of hyperons. Although hyperonization in
dense matter has been studied for several decades (for early studies
see, for example, \cite{Ambartsumyan1960,Glendenning1985,weber_book}),
only during recent years covariant density functional models were
developed which were compatible with the hypernuclear data, mass
and/or radius measurements of neutron stars and the tidal
deformability inferred from the GW170817
event~\cite{Oertel2015,Fortin2016,Providencia2019,Fortin2020,Tolos2017ApJ,Raduta2018MNRAS,Lijj2018b,Li2019ApJ,Lijj2020}.
In this work we use a covariant density functional model whose
parameters are adjusted to available hypernuclear and astrophysical
data and show that both the maximal masses of hypernuclear stars in
both the static and Keplerian limits are incompatible with the mass
range inferred for the light companion in the GW190814 event.  There
are a few alternatives to  the hyperonization scenario discussed
here: (a) deconfined quark matter phases may appear before the
hyperonization threshold; (b) the $\Delta$ resonances may appear in
addition to hyperons. The first effect may have profound implications
for the equation of state and structure of NS and clearly requires a
separate discussion (see \cite{Christian2020ApJ,Pereira2020,Ferreira2020,Blaschke2020Univ,Bauswein2020,Lijj2020}
for recent discussions of various phases and resulting features of
compact objects).
The appearance of $\Delta$ resonances does not affect the maximum
mass of a static NS, but can reduce the radius of the star by
tens of percent~(\cite{Lijj2018b} and references therein). Because our
arguments are based on the static and Keplerian maximum masses of
hypernuclear stars which are very close to those derived for
$\Delta$-admixed hypernuclear matter, our conclusions will not be
affected in this case.

\section{Equation of state of hypernuclear stars}

For our study we used two equations of state of hypernuclear matter
obtained from covariant density functional theory: the first is based
on a functional with the density-dependent meson-baryon couplings with
DDME2 parametrization and its extension to the hypernuclear
sector~\cite{Colucci2013,Dalen2014,Fortin2020,Lijj2018b}.  As an
alternative equation of state, we used the NL3~\cite{Lalazissis1997}
model and its extension to the hyperonic sector.

The Lagrangian density of matter can be written as ${\cal L} = {\cal
  L}_{ {B}}+{\cal L}_l$, where the baryonic contribution is given by
\bea\label{eq:lagrangian_B} {\cal L}_B & = &
\sum_B\bar\psi_B\bigg[\gamma^\mu \left(i\partial_\mu-g_{\omega                                                           
N}\omega_\mu - \frac{1}{2}g_{\rho N}\vectau\cdot\vecrho_\mu\right)
\nonumber\\
&-& (m_B - g_{\sigma N}\sigma)\bigg]\psi_B 
 +  \frac{1}{2} \partial^\mu\sigma\partial_\mu\sigma-\frac{1}{2}
m_\sigma^2\sigma^2 \nonumber\\
&-& \frac{1}{4}\omega^{\mu\nu}\omega_{\mu\nu} + \frac{1}{2}
m_\omega^2\omega^\mu\omega_\mu -
\frac{1}{4}\vecrho^{\mu\nu}\vecrho_{\mu\nu} + \frac{1}{2}
m_\rho^2\vecrho^\mu\cdot\vecrho_\mu, \nonumber
\eea
where $B$ sums over baryons  and $\psi_B$ are the baryonic Dirac
fields with masses $m_B$.  The meson fields $\sigma,\omega_\mu$, and
$\vecrho_\mu$ mediate the interactions among the baryon
fields, $\omega_{\mu\nu}$ and $\vecrho_{\mu\nu}$ represent the field
strength tensors of vector mesons, and $m_{\sigma}$, $m_{\omega}$, and
$m_{\rho}$ are their masses. The baryon-meson coupling constants are
denoted by $g_{iB}$ with $i=\sigma,\omega,\rho$.  The leptonic
contribution is given by
$
{\cal L}_l  = 
 \sum_{\lambda}\bar\psi_\lambda(i\gamma^\mu\partial_\mu -
m_\lambda)\psi_\lambda,
$
where $\lambda$ sums over the leptons $e^-$ and $\mu^-$, which are
treated as free Dirac fields with masses $m_\lambda$.  In the DDME2
model the coupling constants in the nucleonic Lagrangian are
density dependent and are parametrized according to the relation
$g_{iN}(n_B) = g_{iN}(n_{s})h_i(x)$, for $i=\sigma,\omega$, and
$g_{\rho N}(n_B) = g_{\rho N}(n_{0})\exp[-a_\rho(x-1)]$ for the
$\vecrho_\mu$-meson, where $n_B$ is the baryon density, $n_0$ is the
nuclear saturation density, and $x = n_B/n_0$.  This parametrization
has in total eight parameters, which are adjusted to reproduce the
properties of symmetric and asymmetric nuclear matter and the binding
energies, charge radii, and neutron radii of spherical nuclei
(see~\cite{Lijj2018b}).  Its predictions for the main characteristic
parameters for nuclear matter at saturation density are $K=251.15$ MeV
for compressibility, $E_{\rm sym}=32.31$ MeV for symmetry energy,
$L_{\rm sym}=51.27$ MeV for the slope of the symmetry energy.

In the hypernuclear sector, the vector meson-hyperon couplings are
given by the $SU(6)$ flavor symmetric quark model, whereas the scalar
meson-hyperon couplings are determined by fits to empirical
hypernuclear potentials: $U_{\Lambda} = −U_{\Sigma} = -30$ MeV,
$U_{\Xi} = −14$ MeV at nuclear saturation density.  Note that the
Lagrangian of this model has only linear meson-field interaction terms
and the nucleon-meson coupling constants are density dependent.  The
NL3 model is used as an alternative model, which has density
independent meson-nucleon couplings, but contains nonlinear in meson
fields terms; we will comment on the differences between these
models. For illustration, we will also consider hypernuclear stars
which contain a $\Delta$-resonance component, again within the DDME2
model~\cite{Lijj2018b}.

\begin{figure}[tbh] 
\includegraphics[width=\hsize]{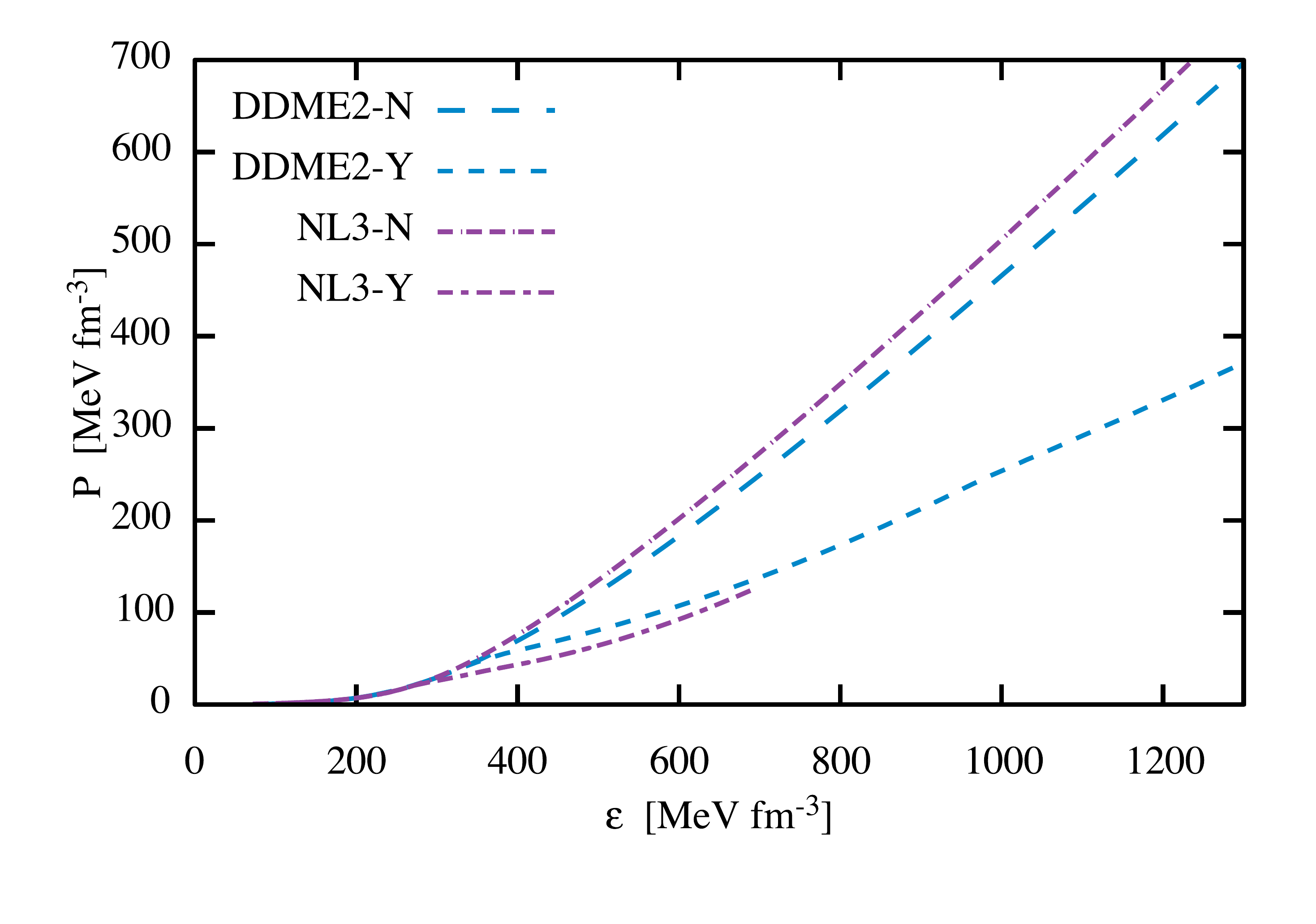}
\caption{ The equation of state of dense matter for two cases: purely
  nucleonic matter (N) and hypernuclear matter (Y) which includes the
  full baryon octet. The long- and short-dashed lines show the results for the
  DDME2 parametrization. The dash-dotted and short-long-dashed lines show the same 
  for  the NL3 parametrization. Note the softening of the equation of
  state (reduction of the pressure) in each case caused by the
  onset of hyperonization. }
\label{fig:EOS}
\end{figure}

The equations of state of purely nucleonic matter and hypernuclear
matter are illustrated in Fig.~\ref{fig:EOS}.
\begin{figure}[tbh] 
\includegraphics[width=\hsize]{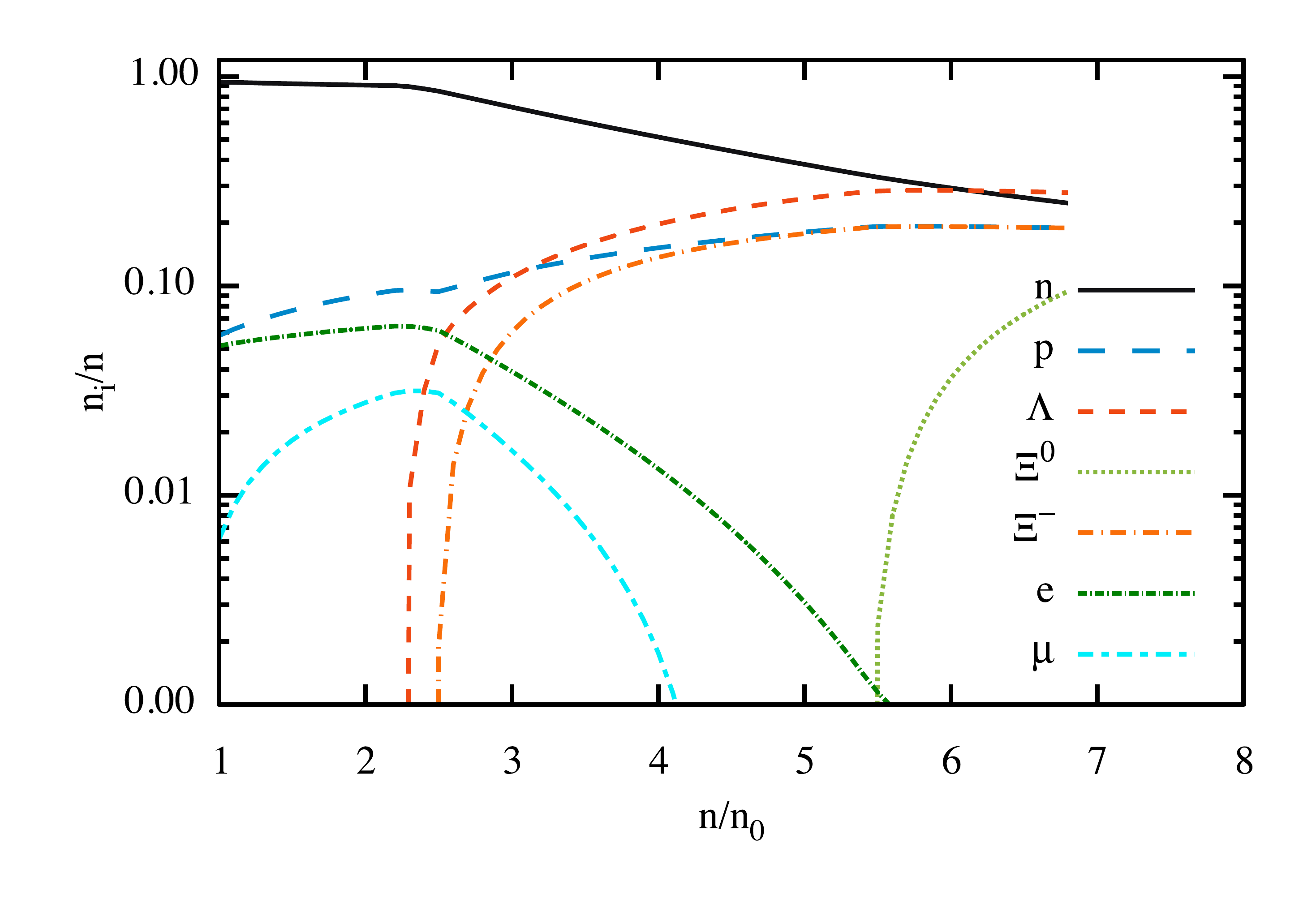}
\caption{ Particle fractions $n_i/n$ ($i$ refers to baryons in the
  baryon octet) in hypernuclear matter according to the DDME2 model. Here
  $n$ denotes baryon density, and $n_0$ is the saturation density of
  ordinary nuclear matter. The results for NL3 model show the same
  features and are not shown here.}
\label{fig:composition}
\end{figure}
The composition of hypernuclear matter computed for the DDME2
parametrization is shown in Fig.~\ref{fig:composition}. It is seen that,
in the relevant density range, the $\Sigma$-hyperons do not appear and that
the dominant hyperonic component is the $\Lambda$-hyperon. 
Among the hyperons, the $\Lambda$-hyperon properties are best
constrained due to the tuning of the interactions (in particular the
coupling to the $\sigma$-meson) to the $\Lambda$ single and
double hypernuclei ~\cite{Dalen2014,Fortin2020}. The variations in the
magnitude of the $\Lambda$-hyperon potential in nuclear matter within
various extensions of the DDME2 model to the hypernuclear sector are
too small to affect the stellar properties and the equation 
of state~\cite{Raduta2018MNRAS,Dalen2014,Lijj2018b,Fortin2020}.

\section{Mass-radius relations of compact stars}

The general relativistic structure equations of compact stars
\cite{Tolman,OV} for the hypernuclear model equations of state shown
in Fig.~\ref{fig:EOS} were solved for spherically symmetric
(nonmagnetized) stars in the cases of static (nonrotating) and
maximally rotating (Keplerian) stars. The rotating configurations were
generated using the public domain RNS code\footnote{\tt
  www.gravity.phys.uwm.edu/rns/}. Note that in both cases the stable
configurations of hypernuclear stars are determined by the
Bardeen-Thorne-Meltzer criterion \cite{BTM_methods}, which implies
that a star is stable only as long as its mass is increasing with the
central density.  Thus, along the rising branch on the mass-radius
curve the stars with the maximum mass for any given sequences are the
last stable configurations.
\begin{figure}[t] 
\includegraphics[width=\hsize]{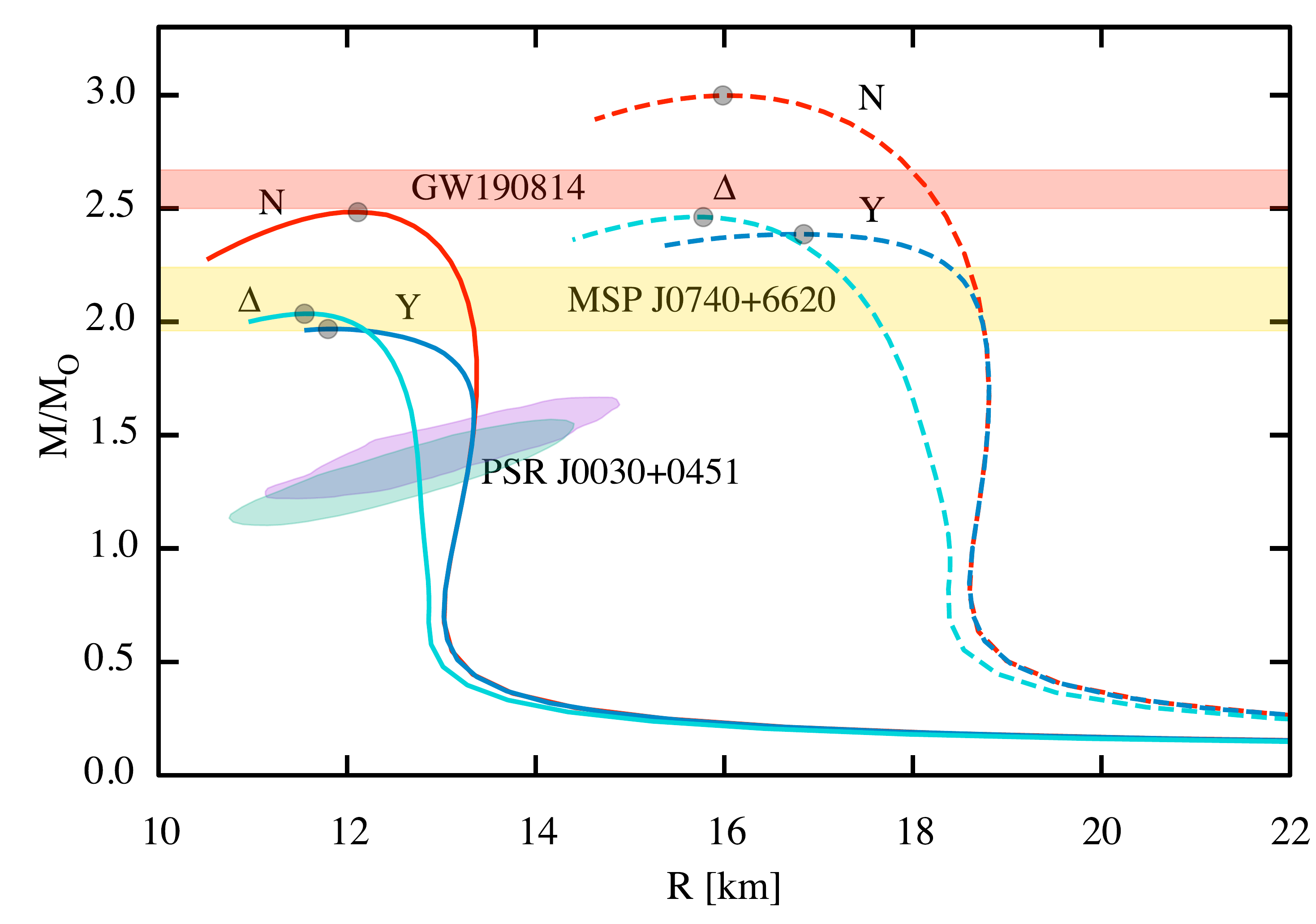}
\caption{The mass-radius relations for nonrotating (solid lines) and
  maximally rotating (dashed lines) nucleonic ($N$), hypernuclear
  ($Y$) and $\Delta$-admixed-hypernuclear ($\Delta$) stars.  The
  colored areas show the constraints inferred from the most massive
  pulsar MSP J0740+6620~\cite{Cromartie:2019kug}, the mass-radius
  limits inferred from the NICER
  experiment~\cite{Miller:2019cac,Riley:2019yda} and the mass limits
  from GW190814~\cite{Abbott:2020khf}. The circles indicate the
  maximum masses of the sequences, to the left of which the stars are
  unstable.}
\label{fig:M_vs_R} 
\end{figure}

In Fig.~\ref{fig:M_vs_R} we show the mass-radius relations of
nucleonic and hypernuclear stars based on the DDME2 parametrization
for static and maximally rotating configurations. Nucleonic
models in the static case reach a maximum mass of $2.48M_{\odot}$
(with a radius of $R=12.1$ km), which makes them marginally compatible with
the NS interpretation of the light companion of GW190814. The
maximally rotating nucleonic models reach masses up to
$3 M_{\odot}$ and thus comfortably account for a NS in GW190814.
 The sequences for the NL3 model show the same behavior whereby the
  maximum mass of nucleonic stars is larger whereas that of
  hypernuclear stars is lower than in the DDME2 model. This implies
  that our arguments are even stronger enforced by the NL3 model. For
  comparison, we also show the sequences of static and  maximally
  rotating $\Delta$-resonance admixed-hypernuclear star with the same 
DDME2 model~\cite{Lijj2018b}; as anticipated the maximum masses in
this case are not modified substantially, but the radii of the models
are smaller than for nucleonic and hypernuclear models.

The nucleonic models satisfy the complementary to GW190814
constraints, specifically, the lower bound on the maximum mass placed
by MSP J0740+6620~\cite{Cromartie:2019kug}, and the mass-radius limits
inferred from the NICER experiment~\cite{Miller:2019cac,Riley:2019yda}
giving, for example, a radius of $R = 13.3$ km for a
$M = 1.33 M_{\odot}$ star. Turning to the hypernuclear models, we note
that the softening of the equation of state triggered by the
hyperonization leads to a lower maximum mass compared to the nucleonic
case. The maximum mass of static hypernuclear stars is
$M\simeq 2.0M_{\odot}$ as is thus consistent with the massive pulsar
MSP J0740+6620, but clearly is inconsistent with a NS in GW190814. The
maximally rotating Keplerian models of hypernuclear stars have maximum
masses $\le 2.3M_{\odot}$. This implies that the maximal rotation is
not sufficient to raise masses of hypernuclear stars to the required
value $2.5M_{\odot}$. Thus, we conclude that independent of the
rotation rate the hypernuclear stars modeled with the DDME2 and NL3
density functionals are incompatible with the light companion of
GW190814 being an NS. While one can attempt to resurrect the
compatibility by modifying the coupling constant for example in the
vector-meson sector by going from $SU(6)$ to $SU(3)$ symmetry (see,
e.g., \cite{JJLi_HF}), we anticipate that the tension between the
theory and the data will still remain.

\begin{figure}[t] 
\includegraphics[width=\hsize]{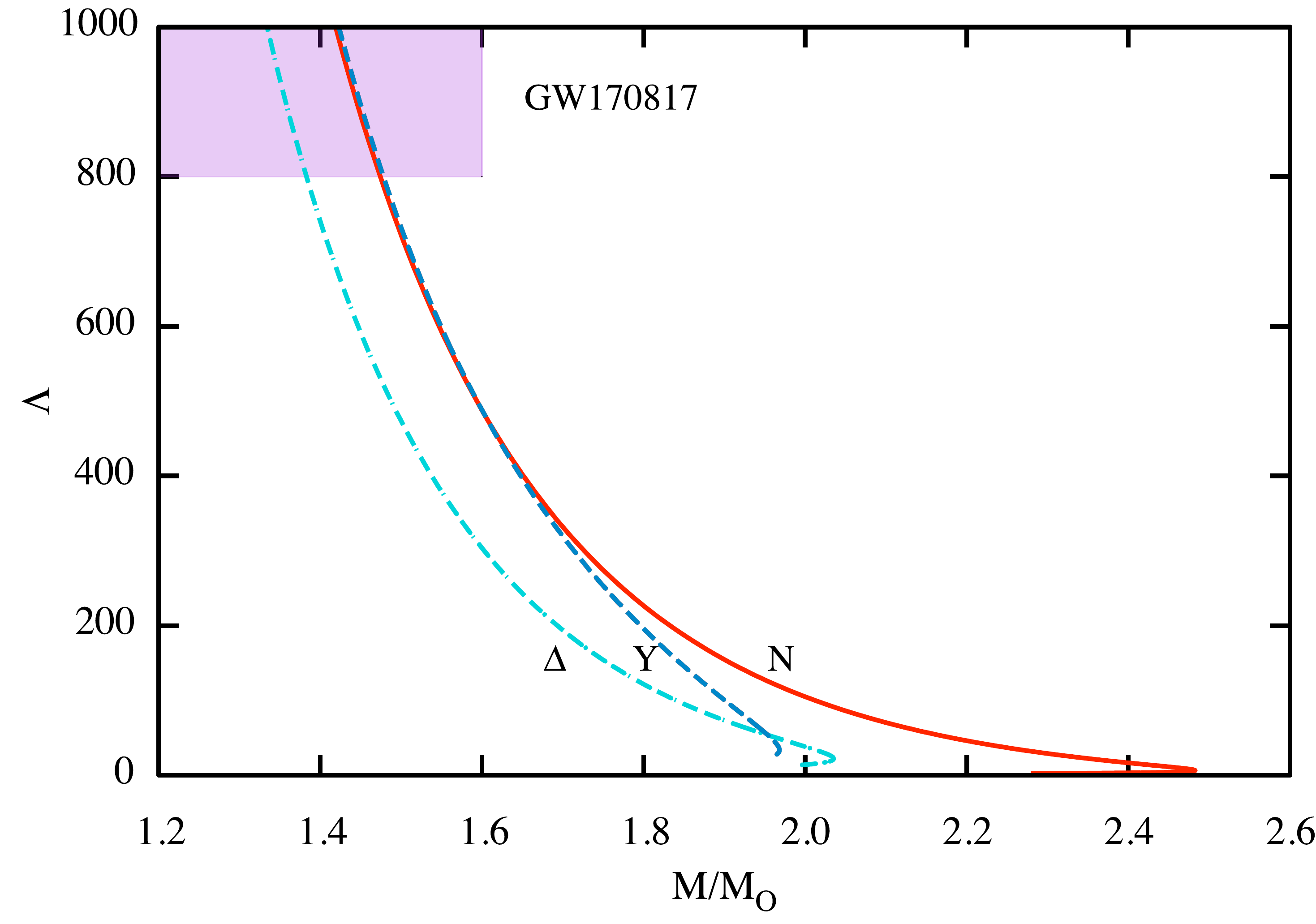}
\caption{The tidal deformability of static nucleonic (solid line),
  hypernuclear (dashed),  and  $\Delta$-admixed-hypernuclear
  stars (dash dotted) according to the DDME2 model. The
  colored area shows the $90\%$ confidence exclusion region for tidal
  deformability of equal mass binary obtained from the analysis of
  GW170817 for the range of NS masses $1.17\le M/M_{\odot}\le 1.6$.}
\label{fig:L_vs_M} 
\end{figure}
Finally, in Fig.~\ref{fig:L_vs_M} we show the tidal deformability vs
mass relations of nucleonic and hypernuclear stars based on the DDME2
parametrization for static configurations along with the $90\%$
confidence exclusion region for equal mass NS-NS binary obtained from
the analysis of GW170817. Our main point here is seen that the massive
stars have very small tidal deformability which is not expected to be
observable at the current level of the sensitivity of gravitational
wave observatories. Note also that the mild discrepancy between the
prediction of the tidal deformability of the DDME2 model for lighter
(canonical) mass NS and the inference from GW170817 can be cured by
accounting for the onset of $\Delta$ resonances~\cite{Lijj2018b}, as
illustrated in Fig.~\ref{fig:L_vs_M}, or adjusting higher-order
expansion parameters of the density functional~\cite{Li2019ApJ}.

\section{Conclusions and outlook}

In this work, we investigated the possibility that the light companion
in the GW190814 event is a hypernuclear compact star. The equation of
state was taken from studies of covariant density functional theory of
hypernuclear matter with the parameters tuned to $\Lambda$
hypernuclear data and astrophysical constraints imposed by the massive
pulsars, NICER experiment, and GW170817 event. We have considered both
static configurations and maximally fast rotating configurations of
hypernuclear stars. As expected, the purely nucleonic stars are
consistent with the involvement of NS in GW190814 even if the star was
nonrotating; adding some degree of rotation would make the nucleonic
models broadly compatible with the scenario involving an NS. However,
such interpretation would imply that matter maintains its purely
nucleonic degrees of freedom up to densities $6.5$ times the nuclear
saturation. A more likely and robust scenario is the hyperonization of
dense matter, in which heavier members of the baryon octet nucleate
once their (in-medium) masses become of the order of the neutron
chemical potential. We found that if hyperonization takes place, then
the maximal masses of sequences of hypernuclear stars are well below
the lower bound inferred for the light companion in GW190814 for
particular density functional studied. The discrepancy can be
mitigated by modifying the coupling constant for example in the
vector-meson sector by going from $SU(6)$ to $SU(3)$ or varying the
$\Xi^-$ potential, but we anticipate that the tension between the
observation and the theory will remain - unless extreme assumptions
are not made about couplings entering the density functionals. The
characteristic parameters of symmetric nuclear matter
(compressibility, symmetry energy, and the slope of the symmetry
energy) at saturation density are well reproduced by the DDME2 model
(this is not the case for NL3 model which has a large slope of the
symmetry energy), therefore the nucleonic sector can be modified only
by adjusting the not-well-constrained parameters corresponding to
higher-order terms in the expansions around the saturation density.

Thus, our main conclusion is that the hyperonization of dense matter
acts strongly against the interpretation of GW190814 involving a
neutron star.  It should be noted that the hypernuclear models are
otherwise broadly consistent with the masses inferred for most massive
pulsars and the mass-radius range for canonical mass neutron stars
inferred by the NICER experiment, see Fig.~\ref{fig:M_vs_R}.
Furthermore, the models we used have been shown to be compatible with
the constraints on tidal deformability of a compact star as inferred
from the GW170817 event, see in particular~\cite{Li2019ApJ}.

We have not addressed in detail the possibility of either phase
transition to quark matter or $\Delta$-resonance admixture in the
hypernuclear matter.  The first issue needs substantial changes in the
input physics, therefore, we do not comment here, rather than refer to
recent studies in
Refs.~\cite{Christian2020ApJ,Pereira2020,Ferreira2020,Blaschke2020Univ,Bauswein2020,Lijj2020}.
Adding $\Delta$-resonances to hypernuclear matter does change the
composition of matter and the radius of the star, but does not
increase (maximum) masses of stellar sequences significantly (see
Fig.~\ref{fig:M_vs_R}).

\begin{acknowledgments}
A.S. is supported by the Deutsche Forschungsgemeinschaft (Grant No. SE
1836/5-1) and European COST Actions ``PHAROS" (CA16214). F.W. is
supported through the U.S.\ National Science Foundation under Grant
PHY-171406.  We are grateful to M.~Alford, D.~Blaschke,
A.~Harutyunyan, M.~Oertel, A.~Raduta, and M.~Sinha for discussions.
\end{acknowledgments}

%

\end{document}